# BAlGaN alloys nearly lattice-matched to AlN for efficient UV LEDs

Logan Williams* and Emmanouil Kioupakis*

Department of Materials Science and Engineering, University of Michigan, Ann Arbor, MI 48109, United States of America

Email: ldwillia@umich.edu; kioup@umich.edu

**Abstract**

The lattice mismatch between AlGaN and AlN substrates limits the design and efficiency of UV LEDs, but it can be mitigated by the co-incorporation of boron. We employ hybrid density functional theory to investigate the thermodynamic, structural, and electronic properties of BAlGaN alloys. We show that BAlGaN can lattice match AlN with band gaps that match AlGaN of the same gallium content. We predict that BAlGaN emits transverse-electric polarized for gallium content of ~45% or more. Our results indicate that BAlGaN alloys are promising materials for higher efficiency UV optoelectronic devices on bulk AlN substrates.

**Main Text**

The efficient generation of UV light is useful in many technological applications, including sensing and measurement of inks and markers, protein analysis and DNA sequencing, UV curing of resins and solvent-free printer ink, and, most importantly, disinfection and sterilization.[1,2]

AlGaN-based LEDs are the most promising semiconducting UV generation technology due to their wide emission spectrum range (from 6.2 eV for AlN to 3.4 eV for GaN), relatively high-efficiency emission of light, ability to be doped n- and p-type, robust mechanical properties, and lack of toxic elements.[2] Current UV generation technology uses mercury-based lamps,[3] but the Minamata Convention on Mercury[4] has over 100 nations pledged to phase out mercury-based technologies.

However, the efficiency of AlGaN-based UV LEDs is still hampered by multiple material issues. Of note, light extraction becomes less efficient for high Al-content LEDs, as the polarization of emitted light switches from transverse electric (TE) (emitted preferentially along the $c$ direction) to transverse magnetic (TM) polarization (emitted preferentially within the $c$-plane) above 68% Al for 3 nm $Al_xGa_{1-x}N$ quantum wells.[5,6] Also, threading dislocation densities (TDD) higher than $10^9$ cm$^{-2}$ are highly detrimental to device efficiency and lifetime.[7,8] The standard, most cost-effective substrate sapphire produces TDDs of $2\times10^9 - 10^{10}$ cm$^{-2}$, while more complex growth processes such as patterned overgrowth or pulsed growth techniques can reduce the TDD to as low as mid-$10^6$ cm$^{-2}$.[8] However, AlGaN grown upon AlN substrates can achieve TDDs as low as $10^3$-$10^4$ cm$^{-2}$,[9] but lattice mismatch between the AlN substrate and subsequent layers (up to 2.4% for GaN) is large enough to nucleate new threading dislocations without proper strain management in device design.

Co-alloying of materials is an established technique that can be applied to vary the band gap and the lattice constants of a material independently of each other. For example, GaAsPBi[10] and GaAsNBi[11] yield alloys lattice-matched to GaAs with a reduced band gap. We have also previously shown that co-alloying produces BInGaN alloys that lattice match GaN with reduced band gaps.[12] Applying Vegard's law to BAlGaN alloys and using our calculated lattice constants of GaN (3.1927Å), AlN (3.1183Å), and w-BN (2.5516Å), we estimate that alloys with a composition of approximately $(B_{0.116}Ga_{0.884})_xAl_{1-x}N$ (B-to-Ga ratio of approximately 1:7.62) are lattice matched to AlN.

Previous theoretical and experimental work on BAlGaN alloys is rather limited. Park and Ahn performed theoretical studies on wurtzite BAlGaN alloys for UV LEDs using multi-band effective mass theory.[13–17] They reported a significant increase in TE-polarized spontaneous emission due to a decrease in lattice mismatch between the quantum well and the AlN substrate, and a decrease in the internal field of the well for low Al content (20%).[13,14] They also reported a decrease in TE-polarized emission with boron addition at high Al content due to a shifting of valence band characteristic from heavy hole to crystal field split-off band.[13] The only experimental exploration of the BAlGaN alloy series was performed by Takano et al.,[18–20] who grew BAlGaN/AlN structures with estimated alloy compositions up to 13% boron using low pressure metalorganic vapor phase epitaxy (1170 °C, 50 mbar) upon SiC substrates. They noted an increase in photoluminescence (PL) with decreased lattice mismatch from AlN, measuring low-temperature PL at 245-250 nm from alloys with 5% boron and room-temperature PL at 260 nm from alloys with 2% boron. However, there has not been a first-principles study of the native properties of BAlGaN alloys.

In this work we explore the thermodynamic, structural, and electronic properties of statistically random quaternary BAlGaN alloys using hybrid-functional DFT. A boron-to-gallium co-alloying ratio of ~1:8 is sufficient to nearly lattice match the alloy to AlN along the *a* direction. Co-alloying with boron and gallium slightly lowers the enthalpy of mixing while greatly increasing configurational entropy. Boron is easily incorporated at the studied concentrations (up to 7.4%). Additionally, replacement of aluminum with boron has negligible effect on the band gap, allowing these alloys to be used in place of AlGaN alloys with an equal fraction of Ga. We predict that BAlGaN alloys maintain favorable TE light polarization for gallium content of 45% or more when grown upon AlN. Our results demonstrate that BAlGaN alloys are promising active-region materials for efficient UV nitride optoelectronic devices.

We performed DFT calculations based on the projector augmented wave (PAW) method,[21,22] as implemented in the Vienna *Ab initio* Simulation Package (VASP).[23–26] The GW-compatible pseudopotentials including 3, 3, 13, and 5 valence electrons were employed for B, Al, Ga, and N, respectively, with a 600 eV plane-wave cutoff. Structural relaxations were performed using the optB86b-vdW functional[27] and a Γ-centered Wisesa-McGill-Mueller Brillouin-zone grid with a minimum period distance of 21.16 Å.[28] Forces on atoms were relaxed to within 1 meV/Å. Band-gap calculations were performed with the functional of Heyd-Scuseria-Ernzerhof (HSE06).[29,30] Random alloys were modeled using special quasi-random structures (SQSs) generated with the Alloy Theoretic Automated Toolkit[31] as 3×3×2 and 3×3×3 wurtzite supercells. Cations were arranged to approximate the pair-correlation functions of random alloys up to 5.125 Å. Five SQSs were generated at each composition for the BAlGaN alloys, and one SQS was generated for each composition for the BAlN, BGaN, and AlGaN alloys used as comparisons for the thermodynamic and electronic properties. Piezoelectric coefficients and spontaneous polarization constants were calculated using PBE with the optB86b-vdW relaxed structures, as hybrid-functional piezoelectricity calculations are computationally too demanding for the alloy supercells. Testing using the binary alloys confirmed that spontaneous polarization results were negligibly affected by the use of the less computationally expensive functional, while our calculated piezoelectric coefficients, although not as accurate as those obtained with hybrid functionals, help make comparisons across materials and understand qualitative trends.

Our thermodynamic analysis shows that the BAlGaN alloys studied are thermodynamically favorable to form at significantly lower temperatures than BGaN and BAlN alloys. The transition temperature between the solid-solution and the miscibility-gap regimes was calculated as a function of composition, $T(x) = \Delta H(x)/S$, where the enthalpy of mixing, $\Delta H$, is the total energy difference between the alloy and the linear combination of the binary compounds that match the composition, and the entropy $S$ is evaluated using the regular solution model, $S = -k_B \sum_{i=1}^{N} x_i \ln x_i$, where $x_i$ is the mole fraction for each of the $N$ alloy ingredients, and $k_B$ is Boltzmann's constant. The transition temperatures (**Figure 1**) for $B_xAl_{1-x}N$ and $B_xGa_{1-x}N$ are well above typical growth temperatures for the materials, which is expected since nitrides are typically grown with epitaxial techniques such as MOCVD or molecular beam epitaxy that occur far from thermodynamic equilibrium. These temperatures cannot be directly correlated to solubility limits, as boron incorporation is limited to ~3% in GaN[32] but 12-15% in AlN[33,34], although their transition temperatures follow similar curves. However, they are useful for comparisons to other B-containing nitride alloys. $B_xAl_{1-9x}Ga_{8x}N$ and $B_xAl_{1-10x}Ga_{9x}N$ both have far lower transition temperatures than $B_xAl_{1-x}N$ and $B_xGa_{1-x}N$ alloys with equal boron content. This is attributable to the much larger configurational entropy of the quaternary alloys, while the enthalpy of mixing is similar. The enthalpy of mixing of the quaternary alloys is only slightly lower than the combined enthalpy of mixing of boron into AlN (or GaN) and enthalpy of mixing of $Al_yGa_{1-y}N$ that the quaternary corresponds to (Figure S1). The reduced enthalpy of mixing and the increased entropy indicate that the studied range of alloys are easy to grow with current growth techniques, in agreement with the previous experimental work.[20]

Our structural calculations show that the addition of boron significantly reduces the lattice mismatch with AlN compared to AlGaN alloys of similar Ga content. $B_{0.11x}Al_{1-1.11x}Ga_xN$ is nearly lattice-matched to AlN along the *a* direction (**Figure 2**). Structural relaxations with the same optB86b-vdW functional for all three materials give *a* lattice constants of 3.1927, 3.1183, and 2.5516 Å for GaN, AlN, and wurtzite BN, respectively. Vegard's law with the theoretical lattice constants predicts an optimal B-to-Ga ratio of approximately 1:7.62 ($B_{0.116x}Al_{1-1.116x}Ga_xN$). The *c* lattice constant is closer to that of the AlGaN alloy with similar Ga content than to AlN. This behavior is the opposite of the trend seen in boron doping of InGaN alloys, in which the *c* lattice direction is more strongly affected by boron incorporation than the *a* lattice direction.[12] The exact cause of this behavior is unclear, but we speculate it could be due to anisotropic lattice distortions of the boron local bonding environments.

Our calculations show that the partial substitution of aluminum with boron in $Ga_xAl_{1-x}N$ alloys does not significantly change the band gap for these low boron concentrations. **Figure 3** shows the calculated band gap vs. Ga fraction $x$ for $Al_{1-x}Ga_xN$, $B_{0.1x}Al_{1-1.1x}Ga_xN$, and $B_{0.11x}Al_{1-1.11x}Ga_xN$ alloys, which includes an interpolated rigid shift to correct for the slight underestimation of the band gaps of the binaries (0.74 eV for BN, 0.52 eV for AlN, and 0.26 eV for GaN) by HSE06. The band gap is direct for all calculated alloys. With the rigid shift included, the band gaps of the calculated BAlGaN, AlGaN alloys show good agreement with a 2-component bowing model:

$$E_{gap} = E_{gap,BN} * x + E_{gap,GaN} * y + E_{gap,AlN} * (1 - x - y) - b_{BN-GaN} * xy$$
$$- b_{BN-AlN} * x(1 - x - y) - b_{GaN-AlN} * y(1 - x - y), \quad (1)$$

with $E_{gap,BN}$=10.803, $E_{gap,GaN}$=3.398, $E_{gap,AlN}$=6.032, $b_{BN-GaN}$=5.295, $b_{BN-AlN}$=5.605, and $b_{GaN-AlN}$=0.277 (all values in eV). Our value for $b_{GaN-AlN}$ is within the relatively wide range of experimental values.[35] Our $b_{BN-AlN}$ value is lower than a prior theoretical report by Shen et al. (b=8.55 eV)[36] due to a difference in the choice of the BN reference gap. However, if we also define the BN gap as the energy difference between the BN states most closely resembling the edge states of AlN (13.9 eV) we obtain a bowing parameter ($b_{BN-AlN}$=8.7 eV) in good agreement with Shen et al.[36] Our $b_{BN-GaN}$ value is higher than calculated by Said et al.[37] and estimated by the experimental review of Kadys et al.,[38] but lower than the range calculated by Jiang et al.[39] Since gallium content controls the band gap in the relevant composition range of these alloys, knowledge of experimental AlGaN band gaps allows for easy selection of BAlGaN alloys with the same band gap, although corrections need to be applied for the band-gap shift due to strain in epitaxially strained layers.

We also investigated the valence-band orbital character and corresponding polarization of the emitted light from BAlGaN alloys. Light extraction is difficult from AlN and high-Al-content AlGaN active layers due to a switch from light being emitted TE-polarized and primarily along the *c* direction to TM-polarized and primarily within the basal plane as the Al content increases.[40,41] This is because as the Al content increases in AlGaN, the crystal field split-off band (composed of N $2p_z$ orbitals oriented along the *c* direction) rises relative to the light- and heavy-hole bands (composed of N $2p_{x,y}$ orbitals oriented within the plane), causing the emission polarization to switch when the crystal field split-off band becomes the top valence band. The relative positions of the bands is a complex function of several factors, including the structural *u* parameter, composition, and strain conditions.[41] See the supplemental information for an analysis of the *u* parameter, composition, and band position dependence. The relative difference between the energy of the light/heavy-hole bands ($E_{LH}$ and $E_{HH}$) and the energy of the crystal field split-off band ($E_{CH}$) at Γ for BAlGaN and AlGaN are shown in **Figure 4**. To calculate $E_{LH}$, $E_{HH}$, and $E_{CH}$ for alloys the Γ-point charge densities of the top three valence bands were projected onto atomic orbitals and $E_{LH}$, $E_{HH}$, and $E_{CH}$ were constructed from the energies of the three topmost valence bands weighted by their relative N $2p_x$, $2p_y$, and $2p_z$ character. $E_{HH}$ - $E_{CH}$ was fit to the data for $Al_{1-x}Ga_xN$, $B_{0.1x}Al_{1-1.1x}Ga_xN$, and $B_{0.11x}Al_{1-1.11x}Ga_xN$ using a bowing equation with both 2- and 3-component bowing terms:

$$E_\Delta = E_{\Delta,BN} * x + E_{\Delta,GaN} * y + E_{\Delta,AlN} * (1 - x - y) - b_{BN-GaN} * xy - b_{BN-AlN} \\ * x(1 - x - y) - b_{GaN-AlN} * y(1 - x - y) + C * xy(1 - x - y), \qquad (2)$$

where $E_{\Delta,BN}$=0.273, $E_{\Delta,GaN}$=0.043, and $E_{\Delta,AlN}$=-0.225 eV. The fitted bowing parameters are $b_{BN-GaN}$=-0.226, $b_{BN-AlN}$=-2.139, $b_{GaN-AlN}$=-0.282, and C=-3.760 (all values in eV). Small quantities of boron can act to substantially raise the heavy-hole band relative to the crystal-field split-off band in unstrained group-III nitrides. We predict that lattice-matched BAlGaN alloys emit TE-polarized up to 55% Al, as opposed to ~40% Al for unstrained AlGaN. This is in good agreement with existing literature using multiband effective mass theory.[13] However, strain and quantum confinement both act to raise the heavy-hole band in $Al_{1-x}Ga_xN$,[42] allowing strained $Al_{1-x}Ga_xN$ quantum wells on AlN substrates to emit TE polarized light up to ~85% Al.[42]

We also calculated the piezoelectric coefficients ($\varepsilon_{33}$ and $\varepsilon_{31}$) and spontaneous polarization ($P_S$) values for the BAlGaN alloy series. The values for AlN, GaN, and BN as calculated using the methods listed in this paper are shown in **Table I**. There is good agreement between our results and previous work on the spontaneous polarization constant, but discrepancies exist for the piezoelectric coefficients for GaN and BN: we find smaller values for GaN and larger values for BN than the previous work by Dreyer *et al.*,[43,44] who used the hybrid HSE functional and likely found more accurate values. The values reported in this work do show strong agreement with previous calculations using a different GGA functional performed by Zoroddu et al.[45] However, calculations with the more accurate HSE functional are prohibitively expensive for the larger alloy systems studied here. Thus, the properties for some alloys and the binary compounds are provided in **Table II** for qualitative comparison. The linear interpolation of binary-alloy properties shows very similar behavior for the studied BAlGaN alloys and AlGaN alloys, shown in Figure S.7. However, the values of both $\varepsilon_{33}$ and $\varepsilon_{31}$ calculated for the BAlGaN alloys are

~0.01-0.03 C/m$^2$ smaller in magnitude than the linear interpolation of their binaries, while P$_S$ is ~0.2-0.4 C/m$^2$ larger. Comparing B$_{0.056}$Al$_{0.44}$Ga$_{0.5}$N and Al$_{0.5}$Ga$_{0.5}$N shows that BAlGaN alloys have larger P$_S$ and smaller magnitude $\varepsilon_{33}$ and $\varepsilon_{31}$ than AlGaN alloys with equal gaps, potentially enabling polarization-engineering designs with BAlGaN alloys. E.g., previous work by Schubert et al. has shown that polarization matching can be used to reduce efficiency droop caused by electron losses in the active layer of InGaN LEDs.[46] However, further work is necessary to fully access the piezoelectric properties of BAlGaN alloys.

In conclusion, we studied the effect of co-alloying boron and gallium into AlN on its thermodynamic, structural, and electronic properties using first-principles calculations. We find that partial replacement of aluminum by single-digit percent boron in AlGaN alloys has negligible effect on the band gap, and causes a large reduction in the *a* lattice constant to match that of AlN. Lattice matching to AlN allows device designs that take advantage of the low TDD (~10$^3$-10$^4$ cm$^{-2}$)[9] of AlN substrates, greatly reducing Shockley-Read-Hall losses and improving the internal quantum efficiency. Our thermodynamics analysis reveals that alloys at the studied compositions can readily be grown, allowing for the growth of BAlGaN layers lattice-matched to AlGaN that emit in the 6.03 – 4.19 eV (206-296 nm) range. Our study of the valence-band character indicates that AlN-lattice-matched BAlGaN alloys are expected to emit TE polarized light at up to ~55% Al. Our results show that co-alloying boron with gallium into AlN yields promising BAlGaN alloys epitaxially matched to bulk AlN substrates that can yield higher-efficiency UV optoelectronic devices.

**Acknowledgements**

This work was supported by the Designing Materials to Revolutionize and Engineer our Future (DMREF) Program under Award No. 1534221, funded by the National Science Foundation. This research used resources of the National Energy Research Scientific Computing (NERSC) Center, a DOE Office of Science User Facility supported under Contract No. DE-AC02-05CH11231.

**Figure Captions**

**Figure 1.** Temperature for the thermodynamic transition between the solid-solution alloy phase and the phase separation to the linear combination of group-III nitride binaries as a function of boron fraction x for $B_xAl_{1-9x}Ga_{8x}N$, $B_xAl_{1-10x}Ga_{9x}N$, $B_xAl_{1-x}N$, and $B_xGa_{1-x}N$.

**Figure 2.** Ratio of the alloy to the AlN lattice constant along the in-plane $a$ (left) and out-of-plane $c$ (right) directions for $B_{0.1x}Al_{1-1.1x}Ga_xN$ (blue points) and $B_{0.11x}Al_{1-1.11x}Ga_xN$ (red points) as a function of gallium fraction x, in comparison to $Al_{1-x}Ga_xN$ alloys (solid orange line). The $a$ lattice parameter is reduced more by the substitution of aluminum for boron than the $c$ lattice parameter is.

**Figure 3.** Calculated gap for $Al_{1-x}Ga_xN$, $B_{0.1x}Al_{1-1.1x}Ga_xN$, and $B_{0.11x}Al_{1-1.11x}Ga_xN$ as a function of gallium fraction x. Values were rigidly shifted by a linear interpolation of the shift required to match GaN and AlN to experiment and to match wurtzite BN to literature calculations.[36]

**Figure 4.** Crystal-field splitting of the topmost valence bands ($E_{HH} - E_{CH}$) at the $\Gamma$ point as a function of gallium fraction for $Al_{1-x}Ga_xN$, $B_{0.1x}Al_{1-1.1x}Ga_xN$, and $B_{0.11x}Al_{1-1.11x}Ga_xN$. For unstrained alloys substitution of aluminum with boron increases the upper band-gap limit for TE-polarized light emission.

**Tables**

**Table I.** Spontaneous polarization ($P_S$), and piezoelectric coefficients for AlN, GaN, and BN calculated using PBE with vdW-relaxed structures and the hexagonal reference structure. Compared to literature values (a) using HSE by Dreyer et al. for BN[43] (with sign error on $P_S$ corrected) and for GaN and AlN,[44] and (b) using GGA (PW91) by Zoroddu et al.[45] Close agreement is achieved for the spontaneous polarization values, and approximate agreement for the piezoelectric coefficients. Values predicted are smaller for GaN and larger for BN than those predicted by Dreyer. Note: the shown piezoelectric constants are the "proper" constants as defined by Dreyer et al.[44]

| | This work (all in C/m²) | | | Literature (all in C/m²) | | |
|---|---|---|---|---|---|---|
| Material | $P_S$ | $\varepsilon_{33}$ | $\varepsilon_{31}$ | $P_S$ | $\varepsilon_{33}$ | $\varepsilon_{31}$ |
| AlN | 1.3367 | 1.494 | -0.592 | 1.312[a] | 1.569[a], 1.50[b] | -0.676[a], -0.62[b] |
| GaN | 1.3307 | 0.616 | -0.364 | 1.351[a] | 1.02[a], 0.67[b] | -0.551[a], -0.37[b] |
| BN | 2.1125 | -1.231 | 0.491 | 2.174[a] | -1.068[a] | 0.282[a] |

**Table II.** Spontaneous polarization ($P_S$) and piezoelectric coefficients for BAlGaN and AlGaN alloys. Standard deviations are included for BAlGaN compositions, where five supercells were used at each composition.

| Material | $P_S$ [C/m²] | $\varepsilon_{33}$ [C/m²] | $\varepsilon_{31}$ [C/m²] |
|---|---|---|---|
| $B_{0.037}Al_{0.296}Ga_{0.667}N$ | 1.330 ± .008 | 1.146 ± 0.009 | -0.496 ± 0.004 |
| $B_{0.056}Al_{0.5}Ga_{0.44}N$ | 1.347 ± .006 | 0.971 ± 0.030 | -0.457 ± 0.007 |
| $B_{0.056}Al_{0.44}Ga_{0.5}N$ | 1.332 ± .010 | 0.925 ± 0.006 | -0.436 ± 0.007 |
| $Al_{0.5}Ga_{0.5}N$ | 1.316 | 1.005 | -0.47754 |

**Figures**

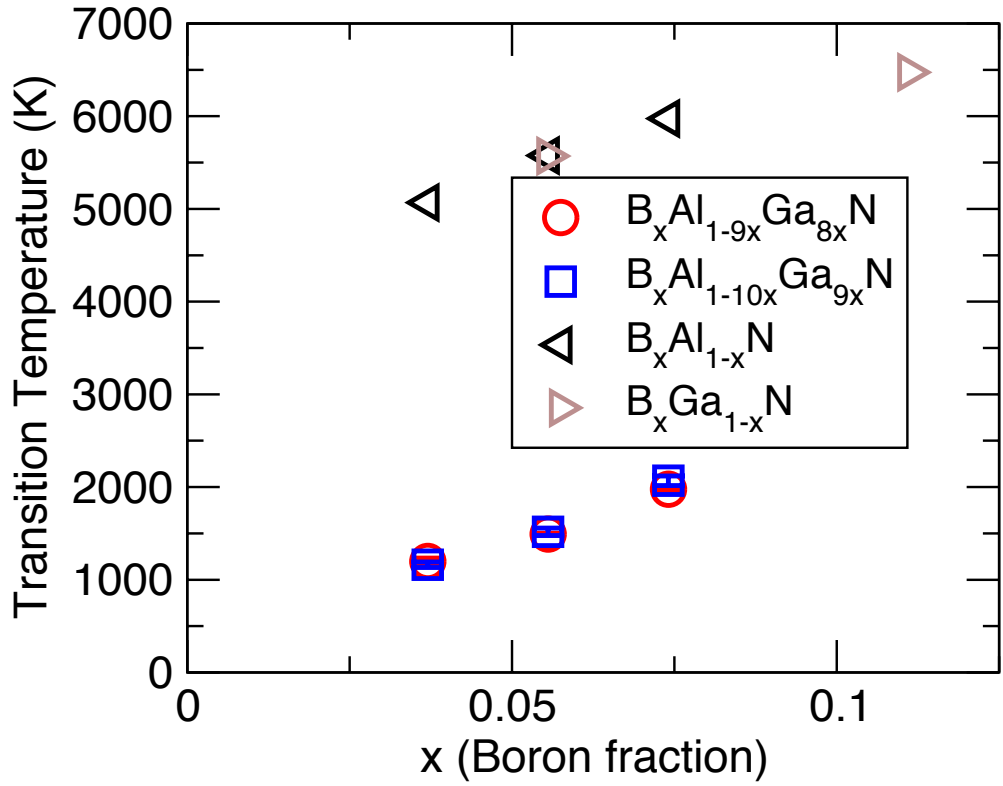

Figure 1

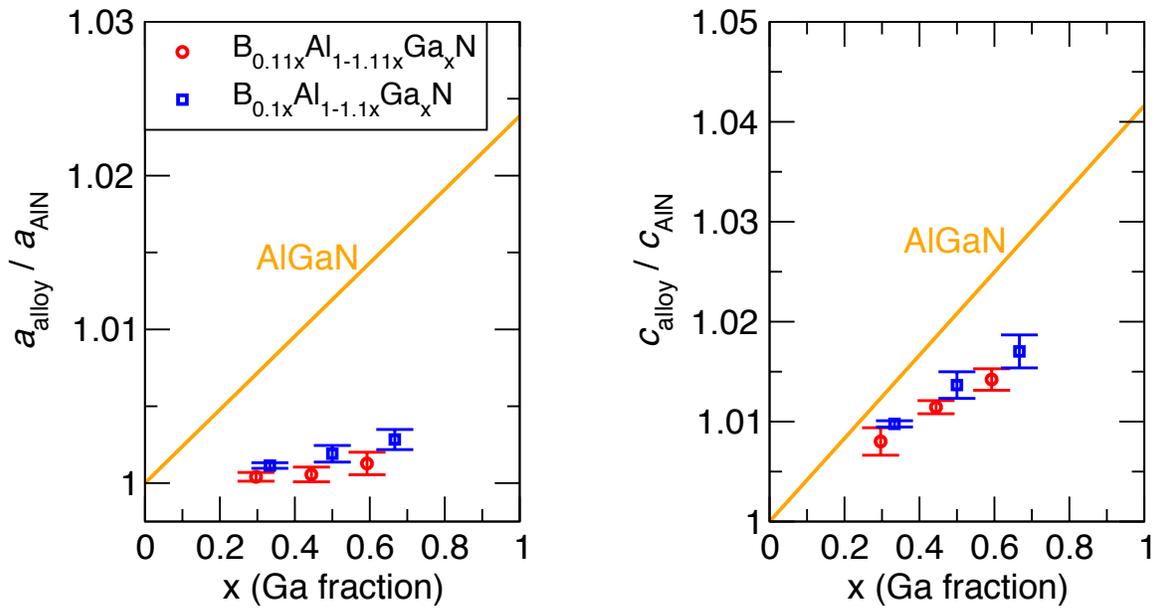

Figure 2

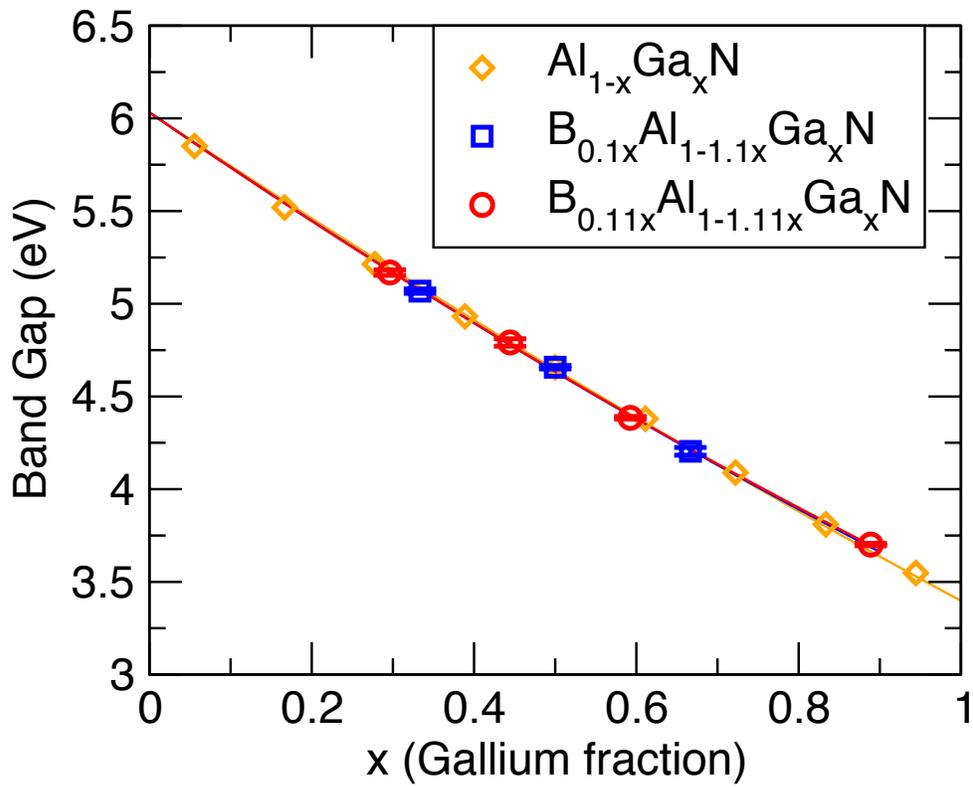

Figure 3

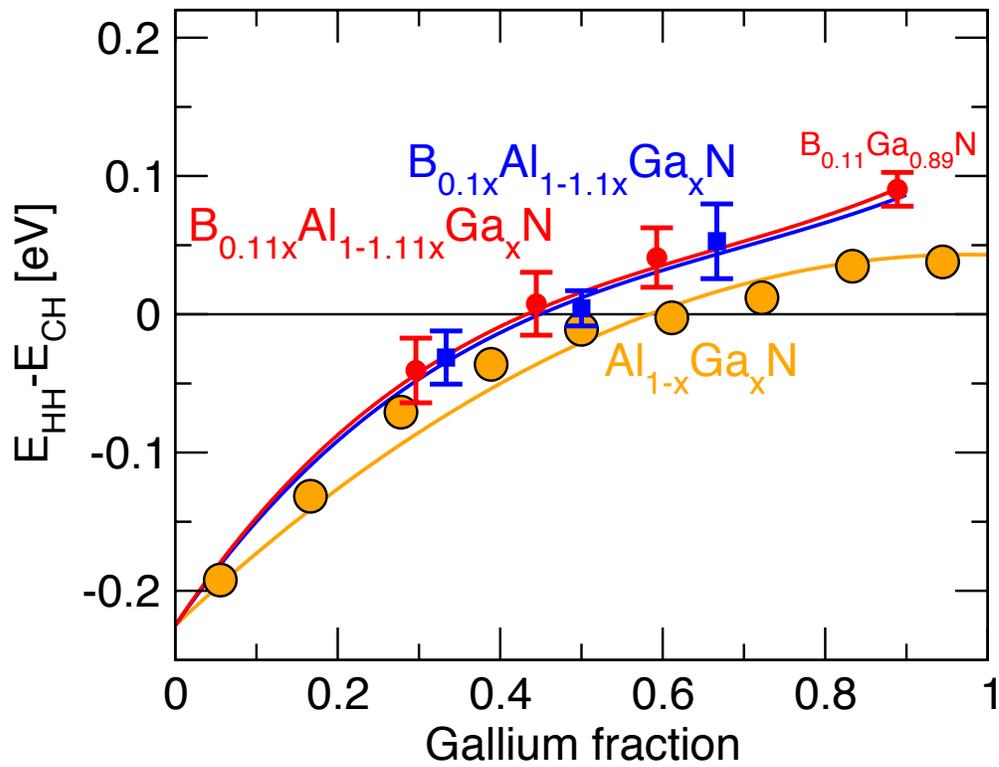

Figure 4

**Supplemental**

In wurtzite group-III nitrides there is a connection between the relative position of their valence bands and the structural $u$ parameter, which quantifies the relative displacement of atoms along the c axis. The ideal tetrahedral wurtzite structure has a $u$ parameter of 0.375, but all group-III nitrides deviate from this value. This effect is partly structural in nature, which can be seen by enforcing different $u$ parameters in the calculations, as shown in Figure S4 along with the natural values for wurtzite BN, GaN, and AlN. In light of this trend, the addition of boron to AlGaN alloys is expected to extend the band-gap range of alloys that emit TE polarized light when grown unstrained. In $Al_{1-x}Ga_xN$, the $u$ parameter does not follow Vegard's law, but is well behaved in a bowing model with b = 0.003, as shown in Figure S.3. For $B_xAl_{1-x}N$, the $u$ parameter follows Vegard's law for low boron content with b nearly zero (-0.000533, Fig. S.5), while the $u$ parameter of $B_xGa_{1-x}N$ (Fig. S.5) has large bowing (0.006) but becomes scattered beyond a few percent boron, where the alloys are not experimentally stable.

The data in Figures S3, S5, and S6 can be combined with the calculations from the $B_xGa_yAl_{1-x-y}N$ series and fit to a multi-component bowing model. If modeled using only two-component bowing terms:

$$u = u_{BN} * x + u_{GaN} * y + u_{AlN} * (1 - x - y) - b_{BN-GaN} * xy - b_{BN-AlN}$$
$$* x(1 - x - y) - b_{GaN-AlN} * y(1 - x - y), \quad (S1)$$

where the bowing parameters are $b_{B-Ga}$ = 0.006278, $b_{Ga-Al}$ = 0.002642, and $b_{B-Al}$ = -0.002892. If including a three-component bowing term C:

$$u = u_{BN} * x + u_{GaN} * y + u_{AlN} * (1 - x - y) - b_{BN-GaN} * xy - b_{BN-AlN}$$
$$* x(1 - x - y) - b_{GaN-AlN} * y(1 - x - y) + C * xy(1 - x - y), \quad (S2)$$

then the bowing parameters are $b_{B-Ga} = 0.006589$, $b_{Ga-Al} = 0.003587$, $b_{B-Al} = -0.000595$, and $C_{B-Al-Ga} = -0.0180253$. As shown in Figure S4, replacement of aluminum by boron in these alloys reduces the $u$ parameter relative to AlGaN alloys with the same band gap (**Error! Reference source not found.**).

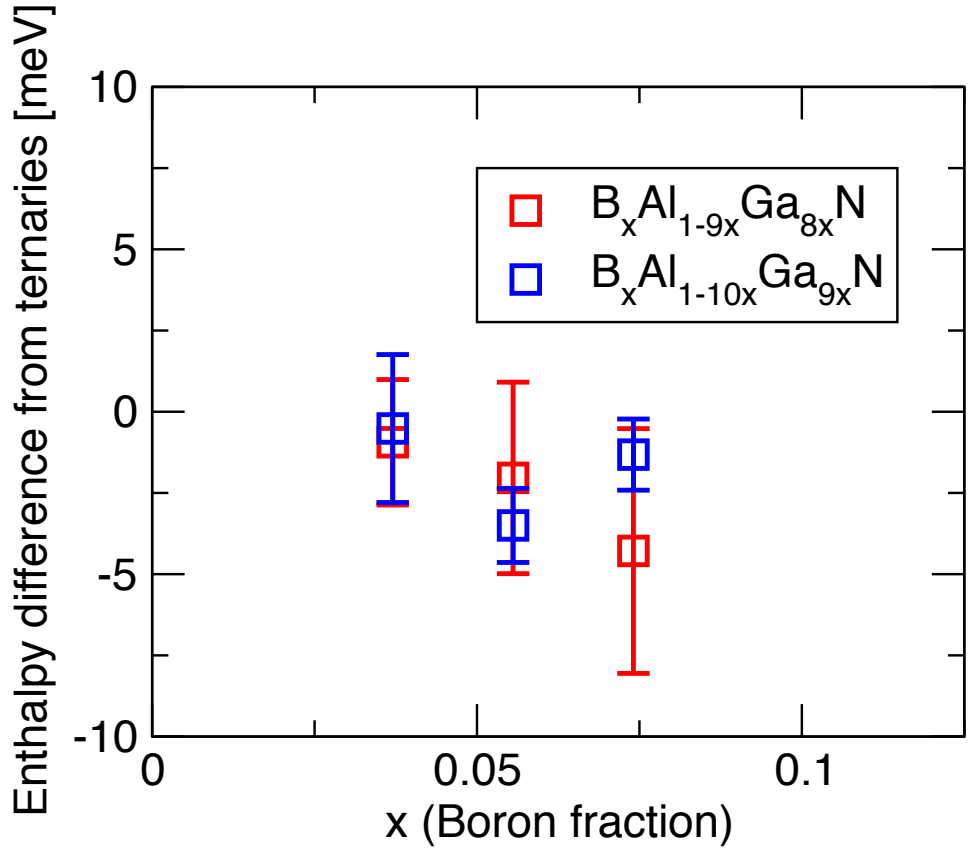

**Figure S.1. The enthalpy difference between ($B_xAl_{1-9x}Ga_{8x}N$, $B_xAl_{1-10x}Ga_{9x}N$) and the sum of the enthalpy for the $B_xAl_{1-x}N$ and $Al_{1-y}Ga_yN$ alloys of the same B/Ga content, respectively.**

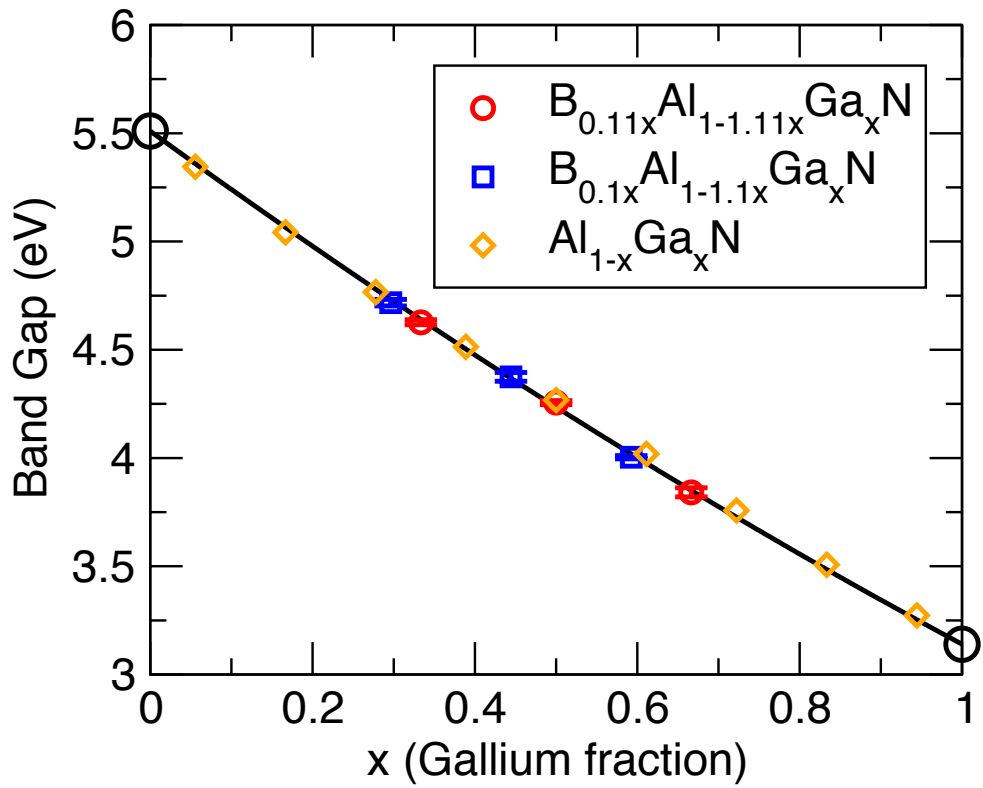

Figure S.2. Band gap as calculated with HSE06 for $Al_{1-x}Ga_xN$, $B_{0.1x}Al_{1-1.1x}Ga_xN$, and $B_{0.11x}Al_{1-1.11x}Ga_xN$ as a function of gallium fraction x.

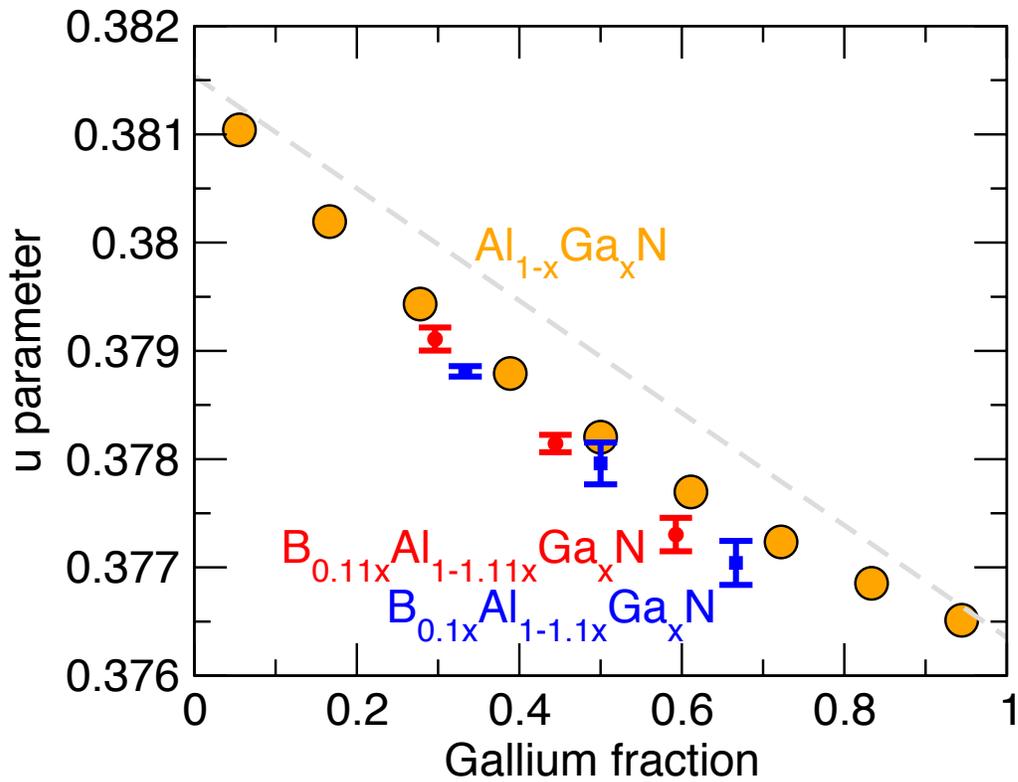

Figure S.3. Structural *u* parameter as a function of the gallium fraction (x) for $Al_{1-x}Ga_xN$, $B_{0.11x}Al_{1-1.11x}Ga_xN$, and $B_{0.1x}Al_{1-1.1x}Ga_xN$ alloys. The u parameter of AlGaN displays a bowing relationship with bowing parameter of b = 0.003. Boron addition reduces the u parameter compared to $Al_{1-x}Ga_xN$ alloys with equal Ga content (i.e. the same band-gap value).

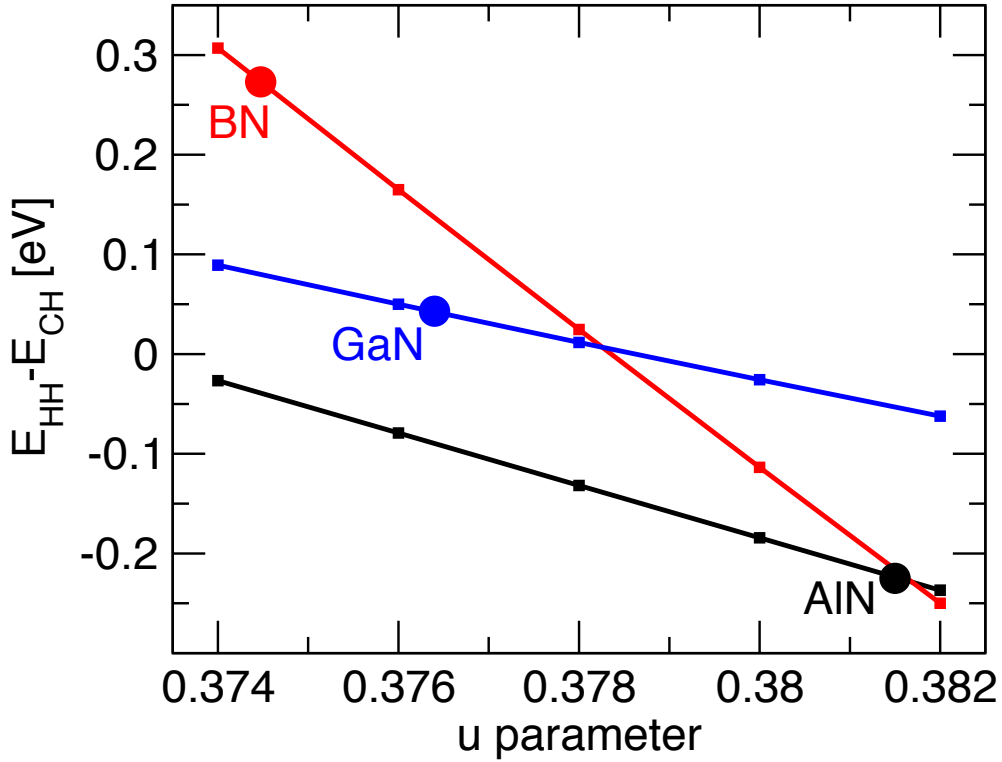

Figure S.4. Crystal Field Splitting ($E_{HH} - E_{CH}$) at the $\Gamma$ point as a function of structural u parameter in wurtzite group III nitrides. Large dots show the natural u parameter and crystal field splitting of the respective materials, which are 0.3745 and 0.2732 eV for BN, 0.3815 and -0.225 eV for AlN, and 0.3764 and 0.043 eV for GaN. All studied materials show a lowering of the heavy-hole band position relative to the crystal field split-off band with an increase in the u parameter.

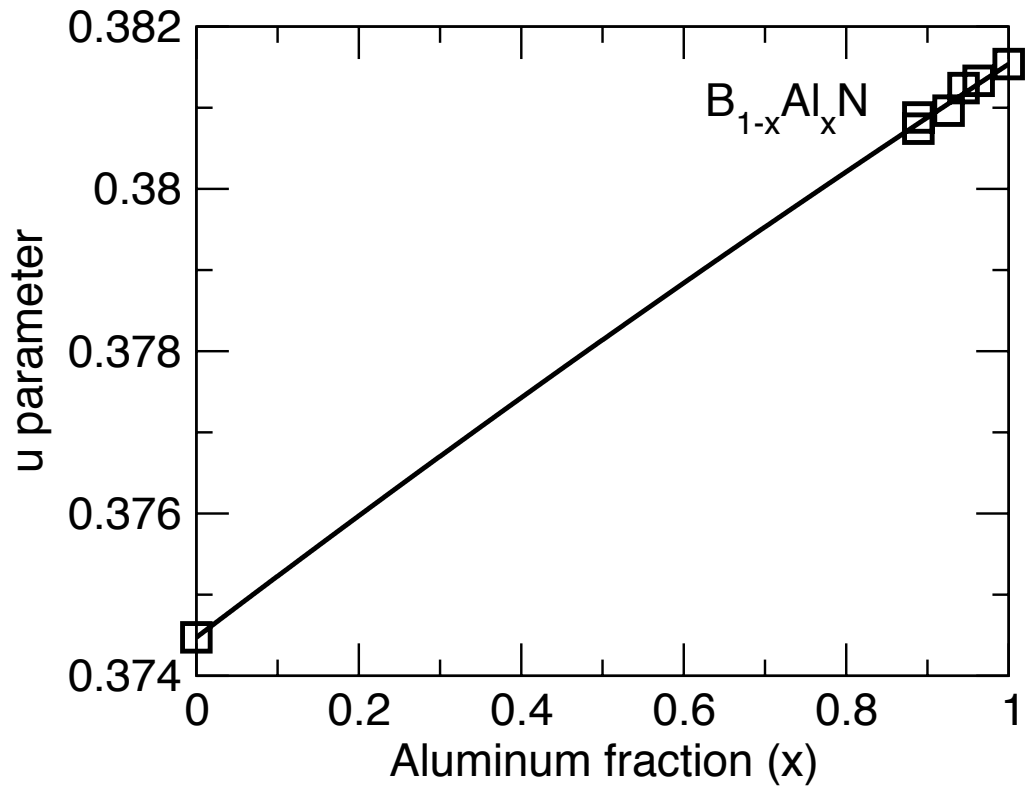

Figure S.5. u parameter vs aluminum content in $B_{1-x}Al_xN$. For low boron content, the u parameter approximately follows Vegard's law. Fitting to a bowing model produces a nearly zero bowing parameter (-0.000533).

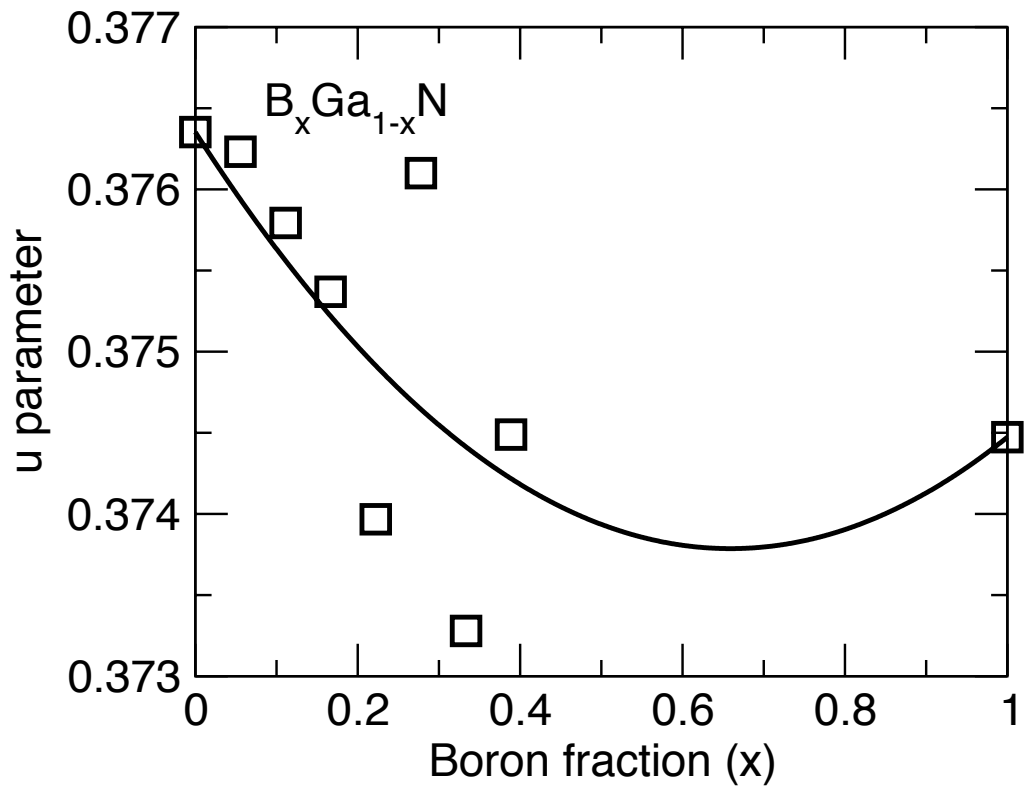

Figure S.6. u parameter vs. the boron content (x) in $B_xGa_{1-x}N$ alloys. The data is scattered at larger boron content, where the alloys are known to be experimentally unstable. The fit in this plot corresponds to b = 0.00591.

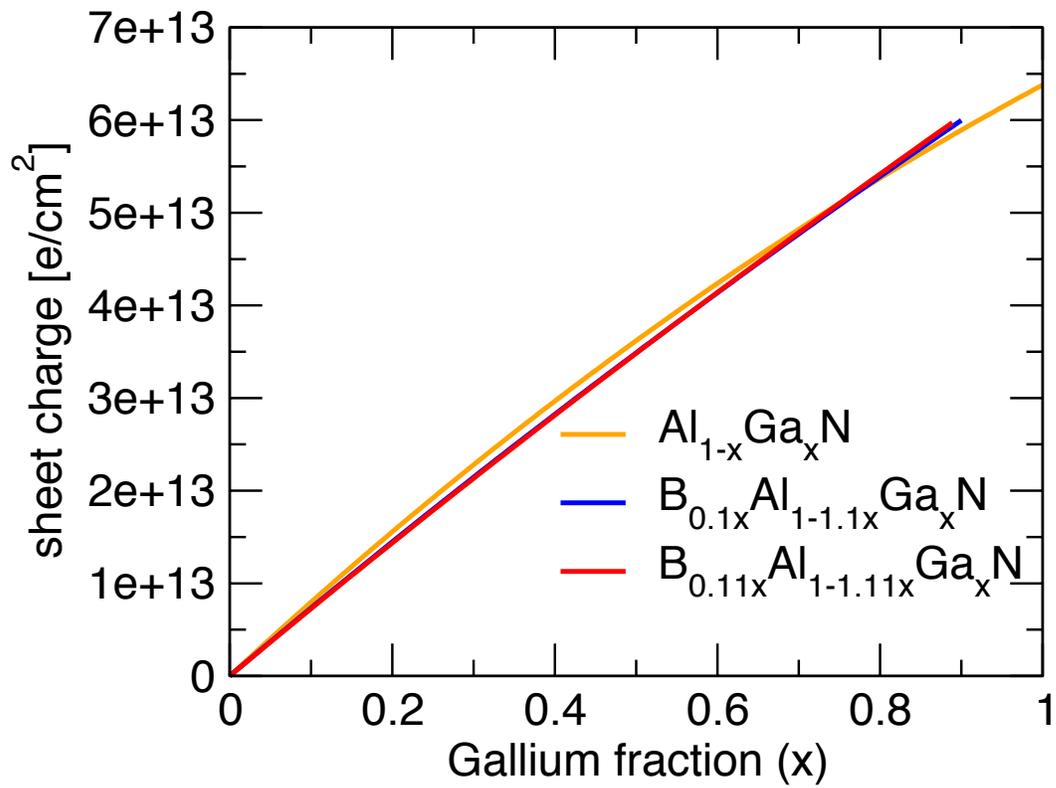

**Figure S.7.** Sheet polarization of an alloy/AlN interface as a function of the gallium content of the alloy. Sheet polarization is calculated using the linear interpolation of binary compound (wz-BN, GaN, AlN) properties.